\documentclass[letterpaper]{article} 
\usepackage{aaai2026}  
\usepackage{times}  
\usepackage{helvet}  
\usepackage{courier}  
\usepackage[hyphens]{url}  
\usepackage{graphicx} 
\usepackage{natbib}  
\usepackage{caption} 
\usepackage{algorithm}
\usepackage{algorithmic}
\usepackage{xcolor}
\newcommand\blfootnote[1]{%
  \begingroup
  \renewcommand\thefootnote{}\footnote{#1}%
  \addtocounter{footnote}{-1}%
  \endgroup
}

\nocopyright
\usepackage{newfloat}
\usepackage{listings}
\DeclareCaptionStyle{ruled}{labelfont=normalfont,labelsep=colon,strut=off} 
\floatstyle{ruled}
\newfloat{listing}{tb}{lst}{}
\floatname{listing}{Listing}
\usepackage{threeparttable}
\usepackage{svg}
\usepackage{xcolor}
\usepackage{utfsym}
\usepackage{amsmath}
\usepackage{multirow}
\usepackage{placeins}
\usepackage{pgffor}
\usepackage{amsmath}
\usepackage{pifont}
\usepackage{fontawesome5}
\usepackage{amsmath}
\usepackage{oplotsymbl}
\usepackage{booktabs}
\usepackage{makecell}
\usepackage{lipsum}
\usepackage{outlines}
\usepackage{minted}
\usepackage{xcolor} 
\definecolor{LightGray}{gray}{0.9}
\newcommand{\answerYes}[1]{\textcolor{blue}{#1}} 
 
\newcommand{\answerNA}[1]{\textcolor{gray}{#1}} 
 
\newcommand{\ours}{\textsc{Tube-Census}}
\newcommand{\faintx}{\textcolor{lightgray}{\ding{55}}}

\usepackage{booktabs}

\usepackage{tablefootnote}

\title{\ours: \\
A Transparent, Replicable, and Large-Scale Census of YouTube
Channels and their Subscriber Counts Over Time}
\author{
    Chloe Eggleston,
    Abram Handler,
    Maria Leonor Pacheco
}
\affiliations{

    University of Colorado Boulder\\
    \{chloe.eggleston, abram.handler, maria.pacheco\}@colorado.edu
}

\begin{document}

\maketitle

\begin{abstract}
YouTube is central to contemporary mass media. 
However, the official YouTube API does not provide access to the full set of creators or creator metadata on the platform.
This lack of basic visibility into the YouTube ecosystem hinders understanding of the platform's creator economy. Researchers currently have no easy, transparent, or replicable way to construct large-scale datasets of YouTube creators and their audiences over time.
This makes it challenging to study vital social questions, such as how changes to the YouTube recommendation algorithm shape creator incentives and by extension the mass media on the platform.
We address this gap with \ours, a large-scale longitudinal dataset of YouTube creators and subscriber counts, constructed by collecting, linking, and organizing nearly two decades of YouTube page captures from the Internet Archive.
This approach is transparent and replicable and does not require interaction with the YouTube API, whose output can change over time.
We validate the coverage of \ours~against prior estimates of YouTube’s size and find that our resource includes creators responsible for at least 30–36\% of all YouTube content. We also find that \ours~provides good coverage of prominent creators. 
To support future research, we hide the substantial complexities of the YouTube identifier system and Internet Archive capture system by distributing our dataset via an easy-to-use \texttt{pip} package.
Finally, we use our resource to complete basic exploratory analysis of YouTube channel content and the mechanisms associated with YouTube channel growth.
\end{abstract}

\section{Introduction}
YouTube creators are core to the platform's attention economy. 
Creators produce videos and audiences subscribe to creator channels to view their content \cite{bergen2022like}.
Observing YouTube channels and their subscriber counts is  therefore essential for research on both the YouTube ecosystem and its impact on global audiences, who collectively watch more than a billion hours of YouTube each day~\cite{Goodrow2017}.
Yet while the YouTube platform is important to many aspects of contemporary social life, there is currently no large-scale public resource listing YouTube channels or their viewership over time.\blfootnote{Software: 
\textcolor{blue}{https://github.com/blast-cu/tubecensus}}

\begin{table}[ht!]
\centering
\small
\begin{tabular}{l ccc ccc}
\toprule
 & \multicolumn{3}{c}{YouTube Channels} & \multicolumn{3}{c}{Subscriber Counts} \\
\cmidrule(lr){2-4} \cmidrule(lr){5-7}
 & API & SB & \textsc{TC} & API & SB & \textsc{TC} \\
\midrule
Transparent  & \ding{51} & \faintx & \ding{51} & \ding{51} & \faintx & \ding{51} \\
Replicable   & \faintx & \faintx & \ding{51} & \faintx & \faintx & \ding{51} \\
Large-scale   & \faintx & \faintx & \ding{51} & \faintx & \faintx & \ding{51} \\
Over time    & \faintx & \faintx & \ding{51} & \faintx & \ding{51} & \ding{51} \\
\bottomrule
\end{tabular}
\caption{Resources for observing what channels are on YouTube and their subscriber counts through time, including the YouTube API (API), Social Blade (SB), and \ours~(\textsc{TC}).}
\end{table}

First, this basic information is not available through the official YouTube API.
YouTube does not expose the set of all channels on the platform and does not provide historical subscriber counts (only current values). 
It also imposes strict daily rate limits. 
These constraints prevent researchers from using the YouTube API to observe creators and audiences on the platform \textit{over time} at a \textit{large scale} (Table 1).

Second, private, for-profit analytics providers partially address these limitations by collecting and reporting historical statistics such as subscriber counts and video views.
However, such historical data are often limited to the top performing channels and limited time windows.
For example, the firm Social Blade lists only the top 5,000 channels at a given time, and provides historical metadata spanning at most the past three years~\cite{SocialBlade}. 
As a result, these services are not suitable for performing large-scale longitudinal analyses of a large set of channels on the platform.
They also often charge fees, lack transparency with respect to their collection procedures, and may discontinue or alter their services without notice. 
These characteristics mean that these private providers are not \textit{transparent} or \textit{replicable} (Table 1), which is required for rigorous research.

Finally, while recent academic work proposes an open and replicable technique for sampling individual YouTube videos~\cite{prefix, dial}, these methods rely on a specific property of the YouTube video ID format and thus can not be used to observe large numbers of channels.
Moreover, the most efficient of these methods requires thousands of requests to the YouTube servers to sample a single video (because it performs random sampling over a massive identifier space).

\begin{figure}[t!]
\begin{minted}
[
framesep=2mm,
baselinestretch=1.0,
bgcolor=white,
fontsize=\small,
]
{python}
from tubecensus import TubeCensus
tc = TubeCensus()
users = tc.sample(1000, by='username')
print(users[0]) 
# >>> {"url": "/user/smosh", ...}
tc.fetch("/user/smosh", closest="201301")
# >>> {"subs": 6561257, ...}
\end{minted}
\caption{The API for the \ours~package.}\label{code1}
\end{figure}

In total, limitations in current resources introduce a fundamental gap in our ability to observe the YouTube creator economy across time.
This has important implications for our understanding of platforms, the mass media ecosystem, and YouTube itself.
Because researchers do not have a basic inventory of the channels on YouTube or their historical subscriber counts,
it is difficult to study fundamental but socially-important questions, such as how creators build audiences on the YouTube platform or how YouTube policies change what people watch online.

To fill this gap, we propose \ours, a large-scale dataset of YouTube channels and their subscriber counts across almost two decades of the YouTube platform.
We construct \ours~by mining the Wayback Machine from the Internet Archive, which periodically crawls and stores copies of websites to preserve their content at specific dates. 
We use this resource to process all captures of YouTube channel URLs through the end of 2023, obtaining more than 106M channels and their historical subscriber data dating back to 2006.
This process requires carefully collecting and resolving different YouTube channel identifiers, which have changed substantially throughout the history of the platform.
Because we built \ours~by mining the Internet Archive, our data processing pipeline is open and transparent, and can be both reviewed and replicated by other researchers.
Moreover, our data collection procedure is performed on static captures of a given timestamp, and this is replicable in comparison to the YouTube API, which changes frequently (e.g., metadata fields).

Our resulting dataset contains over 106M unique channels and their historical subscriber counts. It is roughly 20,000$\times$ larger than the public channel lists available from existing private resources such as Social Blade, and spans the entire lifetime of the platform.

Concretely, \ours~consists of (1) a list of unique channels on YouTube, and their corresponding Wayback Machine archive locations and (2) a \texttt{pip} package to query longitudinal subscriber counts from channels in the census.
Upon publication, we will release our census and companion \texttt{pip} package (Figure~\ref{code1}).

\begin{figure}[h!]
    \centering
    \includegraphics[width=0.9\linewidth]{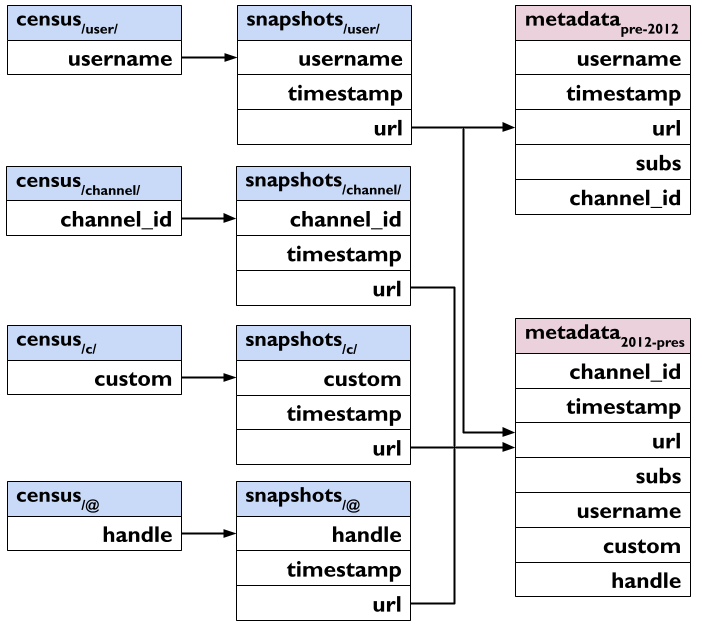}
    \caption{A schema diagram illustrating the data resources we provide. 
    To the left, we have the census: a list of unique YouTube channels, associated with the locations of all their captures on the Wayback Machine. 
    The census abstracts away the substantial complexity of YouTube channel identifiers through time (shown in the diagram) to create a simple list of channels on YouTube.
    To the right, we have the data collected using our \texttt{pip} package, which constitutes longitudinal subscriber counts for channels in the census.
    In this work we use the package to collect, validate, and then release two cohort datasets of subscriber counts for prominent YouTube channels.}
    \label{spec}
\end{figure}

Beyond providing data resources, our work also suggests a path forward for social media research infrastructure in an era of increasing platform API restrictions.
In the early days of social media, platforms allowed researchers to directly observe their data via APIs or the open Web.
More recently, however, many social media platforms have eliminated or significantly restricted free access to their data~\cite{Freelon02102018,doi:10.1177/2056305121988929,Ozkula2023}. These restrictions have sparked debate over whether social media research will be increasingly constrained by limited access to platform data~\cite{Bruns19092019,10.3389/fsoc.2023.1145038}.
By leveraging archival web data from the Internet Archive, our approach provides one path forward for building transparent, replicable, no-cost, and large-scale social media measurement.

\section{Related Work}
\begin{figure*}[ht!]
    \centering
    \includegraphics[width=1\linewidth]{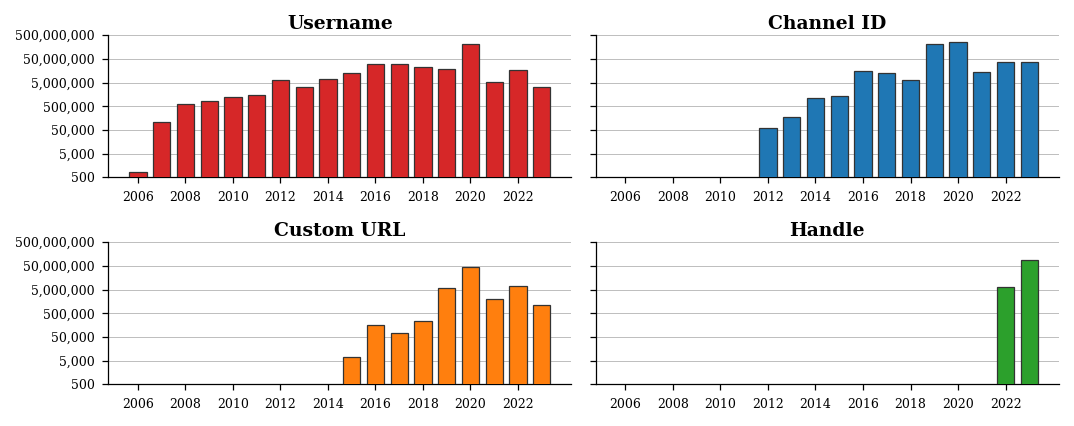}
    \caption{Number of YouTube channel page captures on the Wayback Machine for each URL format per year.}
    \label{bars}
\end{figure*}

\paragraph{Sampling YouTube Data} Sampling YouTube data at scale remains a core challenge for computational social science. The platform does not offer a mechanism for random sampling or bulk retrieval via its public API, forcing researchers to rely on indirect strategies that are often biased or difficult to reproduce. Early approaches to finding YouTube content used opportunistic strategies, such as collecting video URLs shared on Twitter or Reddit~\cite{curtain}, or recursively crawling YouTube's recommendation graph from seed videos~\cite{mislove}. While these strategies can help study diffusion and recommendation dynamics, these methods introduce selection bias: social media samples overrepresent popular, externally visible videos, while recommendation crawling reflects algorithmic dynamics rather than the underlying corpus. Such biases limit statistical representativeness and hinder analysis. 

Another line of research attempts to solve the challenge of discovering YouTube content by exploiting the structure of YouTube's video identifiers, which use Base64-encoded 64-bit IDs and thus enable approximate random sampling by querying random identifiers. 
Researchers have used the prefix search method to estimate the size of the platform using wildcard queries (i.e. partial identifier searches that match any completion of a given prefix)~\cite{prefix}. 
Other variations on this idea exploit the case insensitivity of identifiers to find content, using a technique known as dialing~\cite{dial}. 
These methods approximate random sampling, but are sensitive to API changes and cannot capture historical trends.
They also impose very high computational costs because they require sampling from a very large space of possible identifiers to discover content.

To address these challenges, we use the Internet Archive's Wayback machine to mine archived YouTube channel URLs, following
previous ad-hoc qualitative work in media history that has used the Internet Archive to study YouTube history~\cite{aasman2019traces}.
We extend this informal analysis with a systematic and scalable approach to data collection that allows computational research at scale.

\paragraph{Studies of YouTube} Previous research on YouTube has examined the platform from multiple perspectives, including engagement networks~\cite{Wattenhofer_Wattenhofer_Zhu_2021}, user viewing behavior~\cite{Park_Naaman_Berger_2021}, monetization systems~\cite{curtain,10.1145/3555149}, and ideological information flows~\cite{brown2022echo,10.1145/3351095.3372879}. These areas have gained prominence partly because they are among the few that can be studied with the data YouTube makes accessible to researchers, or that researchers can observe from specialized data collection. For example, system audits of monetization and moderation have used self-reported data from panels of volunteer creators~\cite{10.1145/3555149}, while studies of political and ideological content have typically combined seeded recommendation crawls~\cite{10.1145/3351095.3372879}, real-time data collected through the browser~\cite{brown2022echo}, shares on social media~\cite{lai2022estimating}, and channel-level metadata~\cite{Dinkov2019PredictingTL} to infer ideological orientation and measure exposure patterns. Because these data sources capture only limited slices of the platform, most existing work has focused on questions that can be answered within partial or constructed samples rather than population-level or longitudinal analyses. \ours\; addresses this limitation by providing large-scale historical coverage of YouTube channels and their subscriber counts, opening new avenues to analyze the evolution and engagement of the audience over time.

\section{Data Collection}\label{sec:framework}

In this section, we describe how we construct \ours~by using the YouTube URL format to find and mine the static HTML pages in public data dumps from the Internet Archive.

\subsection{YouTube Channel URL formats} 
\begin{table*}[ht!]
\small
\centering
\setlength{\tabcolsep}{0cm}
\resizebox{\linewidth}{!}{%
\begin{tabular}
{l|c|c|c|c|c|c|c|c|c|c|c|c|c|c|c|c|c|c}
\toprule
Field & 2006 & 2007 & 2008 & 2009 & 2010 & 2011 & 2012 & 2013 & 2014 & 2015 & 2016 & 2017 & 2018 & 2019 & 2020 & 2021 & 2022 & 2023\\
\midrule
Username
&  $ \quad \squadfill \squadfill$ &  $ \squadfill \squadfill \squadfill \squadfill$ &  $ \squadfill \squadfill \squadfill \squadfill$ &  $ \squadfill \squadfill \squadfill \squadfill$ &  $ \squadfill \squadfill \squadfill \squadfill$ &  $ \squadfill \squadfill \squadfill \squadfill$ &  $ \squadfill \squadfill \squadfill \squadfill$ &  $ \squadfill \squadfill \squadfill \squadfill$ &  $ \squadfill \squadfill \squadfill \squadfill$ &  $ \squadfill \squadfill \squadfill \squadfill$ &  $ \squadfill \squadfill \squadfill \squadfill$ &  $ \squadfill \squadfill \squadfill \squadfill$ &  $ \squadfill \squadfill \squadfill \squadfill$ &  $ \squadfill \squadfill \squadfill \squadfill$ &  $ \squadfill \squadfill \squadfill \squadfill$ &  $ \squad \squad \squad \squad$ &  $ \squad \squad \squad \squad$ &  $ \squad \squad \squad \squad$ \\
Ch. ID
&  $ \quad \squad \squad$ &  $ \squad \squad \squadfillhb \squadfillhb$ &  $ \squadfillhb \squadfillhb \squadfillhb \squadfillhb$ &  $ \squadfillhb \squadfillhb \squadfillhb \squadfillhb$ &  $ \squadfillhb \squadfillhb \squadfillhb \squadfillhb$ &  $ \squadfillhb \squadfillhb \squadfillhb \squadfillhb $ &  $ \squadfill \squadfill \squadfill \squadfill$ &  $ \squadfill \squadfill \squadfill \squadfill$ &  $ \squadfill \squadfill \squadfill \squadfill$ &  $ \squadfill \squadfill \squadfill \squadfill$ &  $ \squadfill \squadfill \squadfill \squadfill$ &  $ \squadfill \squadfill \squadfill \squadfill$ &  $ \squadfill \squadfill \squadfill \squadfill$ &  $ \squadfill \squadfill \squadfill \squadfill$ &  $ \squadfill \squadfill \squadfill \squadfill$ &  $ \squadfill \squadfill \squadfill \squadfill$ &  $ \squadfill \squadfill \squadfill \squadfill$ &  $ \squadfill \squadfill \squadfill \squadfill$ \\
Handle
&  $ \quad \squad \squad$ &  $ \squad \squad \squad \squad$ &  $ \squad \squad \squad \squad$ &  $ \squad \squad \squad \squad$ &  $ \squad \squad \squad \squad$ &  $ \squad \squad \squad \squad$ &  $ \squad \squad \squad \squad$ &  $ \squad \squad \squad \squad$ &  $ \squad \squad \squad \squad$ &  $ \squad \squad \squad \squad$ &  $ \squad \squad \squad \squad$ &  $ \squad \squad \squad \squad$ &  $ \squad \squad \squad \squad$ &  $ \squad \squad \squad \squad$ &  $ \squad \squad \squad \squad$ &  $ \squad \squad \squad \squad$ &  $ \squad \squad \squad \squadfillhb $ &  $ \squadfill \squadfill \squadfill \squadfill$ \\
\midrule
Description
&  $ \quad \squadfill \squadfill$ &  $ \squadfill \squadfill \squadfill \squadfill$ &  $ \squadfill \squadfill \squadfill \squadfill$ &  $ \squadfill \squadfill \squadfill \squadfill$ &  $ \squadfill \squadfill \squadfill \squadfill$ &  $ \squadfill \squadfill \squadfill \squadfill$ &  $ \squadfill \squadfill \squadfill \squadfill$ &  $ \squadfill \squadfill \squadfill \squadfill$ &  $ \squadfill \squadfill \squadfill \squadfill$ &  $ \squadfill \squadfill \squadfill \squadfill$ &  $\squadfill \squadfill \squadfill \squadfill$ &  $\squadfill \squadfill \squadfill \squadfill$ &  $ \squadfill \squadfill \squadfill \squadfill$ &  $ \squadfill \squadfill \squadfill \squadfill$ &  $ \squadfill \squadfill \squadfill \squadfill$ &  $ \squadfill \squadfill \squadfill \squadfill$ &  $ \squadfill \squadfill \squadfill \squadfill$ &  $ \squadfill \squadfill \squadfill \squadfill$ \\
Keywords
&  $ \quad \squad \squad$ &  $ \squad \squad \squad \squad$ &  $ \squad \squad \squad \squad$ &  $ \squad \squad \squad \squad$ &  $ \squad \squad \squadfillhb \squadfill$ &  $ \squadfill \squadfill \squadfill \squadfill$ &  $ \squadfill \squadfill \squadfill \squadfill$ &  $ \squadfill \squadfill \squadfill \squadfill$ &  $ \squadfill \squadfill \squadfill \squadfill$ &  $ \squadfill \squadfill \squadfill \squadfill$ &  $\squadfill \squadfill \squadfill \squadfill$ &  $\squadfill \squadfill \squadfill \squadfill$ &  $ \squadfill \squadfill \squadfill \squadfill$ &  $ \squadfill \squadfill \squadfill \squadfill$ &  $ \squadfill \squadfill \squadfill \squadfill$ &  $ \squadfill \squadfill \squadfill \squadfill$ &  $ \squadfill \squadfill \squadfill \squadfill$ &  $ \squadfill \squadfill \squadfill \squadfill$ \\
\midrule
Total Views
&  $ \quad \squadfill \squadfill$ &  $ \squadfill \squadfill \squadfill \squadfill$ &  $ \squadfill \squadfill \squadfill \squadfill$ &  $ \squadfill \squadfill \squadfill \squadfill$ &  $ \squadfill \squadfill \squadfill \squadfill$ &  $ \squadfill \squadfill \squadfill \squadfill$ &  $ \squadfill \squadfill \squadfill \squadfill$ &  $ \squadfill \squadfillhb \squad \squad$ &  $ \squad \squad \squad \squad$ &  $ \squad \squad \squad \squad$ &  $ \squad \squad \squad \squad$ &  $ \squad \squad \squad \squad$ &  $ \squad \squad \squad \squad$ &  $ \squad \squad \squad \squad$ &  $ \squad \squad \squad \squad$ &  $ \squad \squad \squad \squad$ &  $ \squad \squad \squad \squad$ &  $ \squad \squad \squad \squad$ \\
Join Date
&  $ \quad \squadfill \squadfill$ &  $ \squadfill \squadfill \squadfill \squadfill$ &  $ \squadfill \squadfill \squadfill \squadfill$ &  $ \squadfill \squadfill \squadfill \squadfill$ &  $ \squadfill \squadfill \squadfill \squadfill$ &  $ \squadfill \squadfill \squadfill \squadfill$ &  $ \squadfill \squadfill \squadfill \squadfill$ &  $ \squadfill \squadfillhb \squad \squad$ &  $ \squad \squad \squad \squad$ &  $ \squad \squad \squad \squad$ &  $ \squad \squad \squad \squad$ &  $ \squad \squad \squad \squad$ &  $ \squad \squad \squad \squad$ &  $ \squad \squad \squad \squad$ &  $ \squad \squad \squad \squad$ &  $ \squad \squad \squad \squad$ &  $ \squad \squad \squad \squad$ &  $ \squad \squad \squad \squad$ \\
\midrule
Subs.
&  $ \quad \squadfill \squadfill$ &  $ \squadfill \squadfill \squadfill \squadfill$ &  $ \squadfill \squadfill \squadfill \squadfill$ &  $ \squadfill \squadfill \squadfill \squadfill$ &  $ \squadfill \squadfill \squadfill \squadfill$ &  $ \squadfill \squadfill \squadfill \squadfill$ &  $ \squadfill \squadfill \squadfill \squadfill$ &  $ \squadfill \squadfill \squadfill \squadfill$ &  $ \squadfill \squadfill \squadfill \squadfill$ &  $ \squadfill \squadfill \squadfill \squadfill$ &  $\squadfill \squadfill \squadfill \squadfill$ &  $\squadfill \squadfill \squadfill \squadfill$ &  $ \squadfill \squadfill \squadfill \squadfill$ &  $ \squadfill \squadfill \squadfill \squadfill$ &  $ \squadfill \squadfill \squadfill \squadfill$ &  $ \squadfill \squadfill \squadfill \squadfill$ &  $ \squadfill \squadfill \squadfill \squadfill$ &  $ \squadfill \squadfill \squadfill \squadfill$ \\
\bottomrule

\end{tabular}}
    \caption{Metadata in the HTML pages captured per quarter year. 
    $\squadfill$ (full coverage), $\squadfillhb$ (partial coverage), $\squad$ (full absence). 
    }
    \label{metadata}
\end{table*}

YouTube channels have been represented with four different formats throughout the platform's history.
Our collection procedure considers formats starting from 2006, one year after the platform's founding. 
During this early period, the platform represented channels using case-insensitive alphanumeric usernames of at most 20 characters.
This changed in 2012, when YouTube began to represent channels using 128-bit IDs encoded in base64url. 
By 2015, YouTube added an integration with Google+, so YouTube channels could add custom URLs that linked to Google+ pages. 
These custom URLs were replaced by optional YouTube handles in 2022, which later became mandatory for channels on the platform.
Each of these four formats reference the same YouTube HTML page, which has a unique channel identifier embedded in its metadata.
We use these unchanging channel identifiers to map each alternate URL format to the same channel through time. 

In Table \ref{formats} and Figure \ref{bars}, we show these different URL formats, their corresponding platform time periods, and the number of captures they represent on the Wayback Machine over time.

\begin{table}[h!]
\setlength{\tabcolsep}{2.5pt}
\renewcommand{\arraystretch}{0.9}

\small
\centering
\begin{tabular}{l  l  l  l }
\toprule
Format & Prefix & RegEx & Time \\
\midrule
Username & {\scriptsize \tt /user/} & \scriptsize{\verb|[A-Z0-9]{1,20}|} & '06-'14  \\
\quad Legacy & {\scriptsize \tt /profile?user=} &   \scriptsize{\verb|[A-Z0-9]{1,20}|} & '05-'06\\
\quad Vanity & {\scriptsize \tt /} & \scriptsize{\verb|[A-Z0-9]{1,20}|} & '07-'13\\
Channel ID & {\scriptsize \tt /channel/UC} & \scriptsize{\verb|[A-Za-z0-9_-]{22}|}& '12-\text{Pres}\\
Custom Name & {\scriptsize \tt /c/} &  \scriptsize{\verb|[A-Z0-9]+|}  & '15-'21\\ 
Handle & {\scriptsize \tt /@} & \scriptsize{\verb|[A-Z0-9-_]{3,30}|}  & '22- \text{Pres}\\ 
\bottomrule
\end{tabular}
\caption{YouTube channel URL formats.}
\label{formats}
\end{table}

\subsection{Collecting Wayback Machine Snapshots}

We exploit the structure of the different URL formats along with the Internet Archive's Wayback Machine to create our dataset. 
The Wayback Machine periodically captures and stores HTML snapshots of URLs at specific points in time, preserving the state of each page at the time it was archived.
The Internet archive distributes these captures through an index of timestamps, URLs, status codes, and other associated metadata. 
We use the URL format prefixes described in Table~\ref{formats} to collect and download all successful YouTube channel captures until the end of 2023. Each capture (URL, timestamp) is guaranteed to contain a subscriber count value. 
Therefore, we refer to these tuples as a \textit{channel snapshot}. 
Table~\ref{init_table} shows the number of channel snapshots that we were able to collect for each URL format.\\

\begin{table}[h!]
    \centering
    \begin{tabular}{lcc}
    \toprule
         Format & Channel Snapshots & Unique Channels  \\
         \midrule
         Username & 372M & 34.8M\\
         ID & 683M & 106M\\
         Custom & 62.4M & 5.9M\\
         Handle & 94.5M & 25.4M\\
         \midrule
         Total & 1.21B & At least 106M \\
         \bottomrule
    \end{tabular}
    \caption{Count of total webpage snapshots and channels.}
    \label{init_table}
\end{table}

Because username-based URL formats did not include an explicit channel identifier prior to 2012, we cannot reliably map URLs across formats during this period to de-duplicate channels without querying the YouTube API for each case. Nevertheless, we find that the largest of the four snapshot categories (the channel ID format) contains slightly more than 100 million unique channels. This provides a conservative lower bound on the total number of channels in our dataset, since the remaining formats may include additional channels not captured in the ID category.

\subsection{Metadata for Longitudinal Measurement}

YouTube has repeatedly changed the layout of its website over the course of its development.
These redesigns alter the metadata embedded in the HTML pages archived by the Wayback Machine, which in turn constrains which channel properties can be reliably observed in any given time period.
For example, a channel's total view count and join date were visible on the primary channel page until 2013, after which they were moved to the {\tt /about} subpage. As a result, these fields are not consistently present across all archived captures for a given channel.  

Table \ref{metadata} summarizes the metadata fields available on the primary channel page over time. Two of these fields are particularly stable and important; channel identifiers (username, channel ID, and handle), which are often encoded in HTML metadata tags, and subscriber counts, which are typically displayed near the channel name or integrated into the subscribe button.

Together, these fields form the basis for our longitudinal measurements. References between channel URL formats enable us to map captures across formats without relying on the YouTube Data API, allowing us to combine snapshots from multiple page types to construct higher-resolution time series while remaining unencumbered by API rate limits and data redistribution restrictions. In turn, the subscriber count field provides a historical measure of a channel's audience size at specific points in time, and thus enables long-term longitudinal analysis of channel behavior over the seventeen year period. Crucially, because no YouTube API endpoint exposes historical subscriber counts, these longitudinal measurements were previously infeasible.

\subsection{FAIR Principles}\label{sec:fair}
We adhere to the FAIR principles for scientific research \cite{Wilkinson2016} when designing, collecting and distributing \ours.

\paragraph{Findable } We distribute our dataset on Zenodo, which provides a permanent digital object identifier ({\tt 10.5281/zenodo.18267682}). As of publication, we will distribute our code on GitHub, as well as a package on PyPI.

\paragraph{Accessible} We provide all dataset fragments as comma-delimited tables (CSVs) or JSONLines files, both being primary data formats for tabular data. In addition, since we are uploading our software to PyPI, it can be downloaded using the pip package manager, the standard for Python libraries. 

\paragraph{Interoperable} The CSV format and the JSONLines format are supported by the majority of database systems and data analysis libraries.

\paragraph{Reusable} We opt to redistribute both the census (list of channels and when they are captured by the Wayback Machine) and the cohort channel datasets of subscriber count over time collected with our \texttt{pip} package. We redistribute no additional metadata such as channel descriptions to minimize the risk of revealing personally identifiable information (PII), distributing only the information found in the URL formats (channel IDs / usernames) and their subscriber counts at a given timestamp. Additionally, if a YouTube channel owner requests removal from our dataset, we will honor the request by submitting a deletion request to Zenodo and issuing an updated release that tombstones the corresponding record.

All data in \ours{} are collected from publicly accessible webpages archived by the Internet Archive's Wayback Machine. We adhere to methods that rely solely on the Internet Archive and avoid any interaction with YouTube's servers for both reproducibility and shareability, as data collected in that manner would be subject to their terms of use. This does, however, lead to limitations in data collection. 

\subsection{Limitations}\label{sec:limitations}
\paragraph{Video Linking} In principle, we might augment \ours~by linking YouTube channels with videos, which are also substantially represented in the Internet Archive. While this direction is technically plausible, it would require us to map videos to channel IDs, which cannot be easily achieved without querying the YouTube Data API (or alternative services such as Innertube, YouTube's internal undocumented API that researchers often reverse-engineer to access more granular YouTube data without rate limits). Thus, in order to ensure reusability, this is out of scope for our work.

\paragraph{Collection bias} Our data is inherently biased by the data collection decisions of the collections present in the Wayback machine. Some of these collections, such as web crawls like Common Crawl, have transparent data collection practices whose bias could be quantified.
However, in some cases biases are opaque. 
Researchers who use \ours~would need to be aware of these limitations, and adjust their analysis based on nuances of data collection.
For example, researchers may wish to compute a mean daily growth rate to account for the fact that different channels are captured at different rates (see next section).

\subsection{Potential Negative Social Impacts and Misuse}\label{sec:misuse}
Although \ours{} includes only channel identifiers, these IDs could be used by third parties to link channels to additional platform data and metadata via the YouTube API. As a result, our resource could be misused for unfair profiling, targeted harassment, or other forms of harm towards channel creators. To reduce these risks, we limit our release to the minimal information required for transparency and longitudinal measurement. We encourage downstream users to follow appropriate ethical review and harm-mitigation practices when using or augmenting these data.

\section{Assessing the coverage of \ours}\label{sec:validation}

\ours~proposes that Internet Archive captures are a good way to observe YouTube content because they are transparent, replicable, and can be analyzed without any interaction with the YouTube API.
Nevertheless, because we do not fully understand why the Internet Archive captures particular channels at particular times, relying on Internet Archive introduces questions about the channel coverage in \ours.

\begin{figure}[t!]
    \centering
    \includegraphics[width=\linewidth]{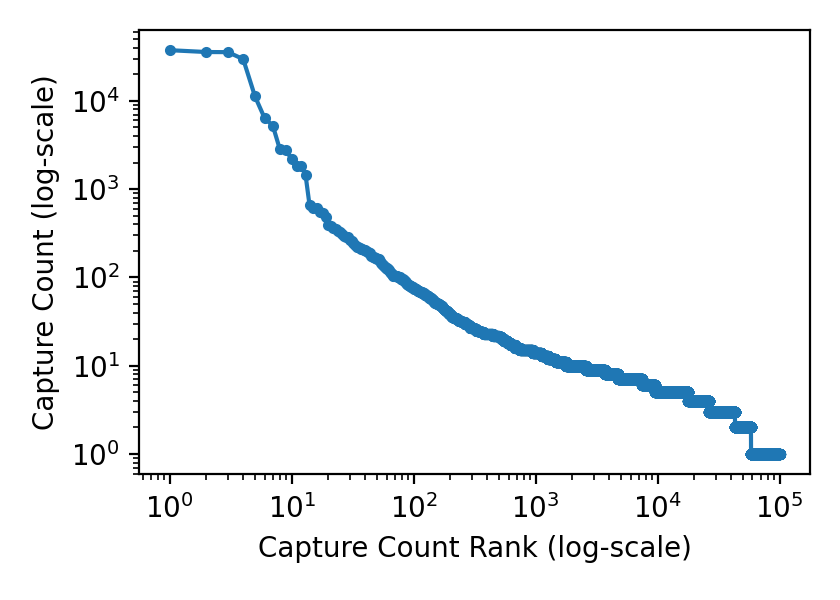}
    \caption{Capture count rank-frequency plot of a random sample of 100,000 channel IDs.}
    \label{longtail}
\end{figure}

We investigate these questions by observing empirical regularities in how the Internet Archive captures YouTube content, as collected in \ours.
In general, we find that the empirical distribution of channel captures in \ours~ follows a long tail (Figure \ref{longtail}).
A small number of channels are captured frequently on the Internet Archive while a long tail of larger channels is captured less frequently.

We therefore analyze the \textit{depth} and the \textit{breadth} of the coverage in \ours.
We find that \ours~likely provides \textbf{deep} coverage of prominent content creators (Sec. \ref{s:creatorcoverage}); the temporal granularity of observations increases for channels with more subscribers who lie at the head of the empirical distribution.
This represents an improvement over current resources for observing YouTube, which only offer subscriber counts for the very largest channels on the platform.
We further find that \ours~provides \textbf{wide} coverage of creators on YouTube.
Many creators from the long tail of the distribution (Figure \ref{longtail}) appear in \ours~at least one time.
The collection of creators in \ours~create  $\approx$ 30-36\% of the videos on the platform (Sec. \ref{cov}).
\ours~offers an index into such content, which is currently unobservable.

\subsection{Coverage of Content Creators}\label{s:creatorcoverage}

To gauge how well \ours~covers creators on YouTube, we calculate the rank correlation between a channel's most-recent subscriber count and the number of occurrences of its primary channel URL in our dataset, limiting our calculation to channels that appear at least 20 times in our snapshots. The results are shown in Table~\ref{corr_table}.

In general, we observe moderate correlations between the number of channel snapshots and the number of channel subscribers, across all different URL formats.
This suggests that channels with large audiences are more likely to appear in our dataset, meaning our dataset likely achieves high coverage of creators who are most visible on the platform.

\begin{table}[h!]
    \centering
    \begin{tabular}{lcc}
    \toprule
         Format & Rank Corr. & Count $\geq$ 20  \\
         \midrule
         Username & $\rho=0.354$ & $n=684,763 $\\
         ID & $\rho=0.432$ & $n=577,624 $\\
         Custom & $\rho=0.221$ & $n=284,306$\\
         Handle & $\rho=0.400$ & $n=90,160$\\
        \bottomrule
    \end{tabular}
    \caption{Spearman's Rank Correlation between the number of  channel snapshots and channel subscriber counts.}
    \label{corr_table}
\end{table}

\subsection{Coverage of Videos} \label{cov}
To estimate the number of videos from creators in \ours~as a fraction of the total number of videos on YouTube, we randomly sample channels from our set of 106M channel IDs and collect the number of videos per channel as of October 2025.

However, many YouTube channels have few subscribers, and are captured less frequently on the Internet Archive (Table \ref{corr_table}).
These channels also appear less frequently in our sets of captures.
Therefore, to ensure we draw channels with different subscriber counts, we use stratified sampling to randomly select channels with different numbers of captures in our dataset.
Concretely, we draw 34,000 channels by sampling 2000 channels from among channels with captures greater than or equal to $2^i$ and less than $2^{i+1}$  for $i=0$ to $i=16$. 

We observe the number of videos for each channel and use the proportion of videos in each strata in our dataset to construct a weighted average of video counts. 
This yields a mean of 46.3 videos per channel, with a standard error of 2.3 videos (the square root of the variance across strata). 
Multiplying by 106M known IDs, we form a 95\% confidence interval of (4.39, 5.34) billion videos covered. 
We then compare this estimate with prior estimates of the number of videos on YouTube~\cite{dial,zheng2023tubestats}, and find that we obtain substantial coverage ($\approx$ 30-36\%). 

\section{Querying and Validating \ours}\label{sec:data}

\ours~consists of a large-scale list of creators on YouTube, which can be queried via a \texttt{pip} package to obtain longitudinal subscriber counts. 
We validate this functionality by using our \texttt{pip} package to construct two channel cohort datasets spanning two time periods. 
We then validate each cohort by checking its representation of highly influential channels against external lists of influential channels from Social Blade.
We find that high-subscriber channels in our queried cohorts largely match high-subscriber channels from the commercial platform, validating our results.\footnote{
Private analytics providers do not use transparent or replicable methods and charge money, as described in the Introduction. We only use their top-channels lists to validate our free and open package.
}
We also release these cohort datasets as part of \ours, as constructing these resources required millions of calls to the Internet Archive (which do not need to be repeated by each future research team).

\subsection{Cohort 1 (2006-2013)}\label{sec:data_1}
\subsubsection{Background and rationale}
Our first cohort covers the early years of YouTube.
We observe few captures during this time period relative to dataset as a whole (Figure~\ref{bars}). 
To build our cohort, we therefore observe all known captures during this time period and discretize Internet Archive timestamps to order all channels by the number of unique quarters which contain one capture for the channel.
We then proceed down the ranked list and collect one capture per quarter until we have $\sim$ 3 million captures.
Because the capture frequency is strongly correlated with subscriber count (Table~\ref{corr_table}), this procedure collects captures and subscriber counts for prominent creators.

\subsubsection{Details}
We collect URLs that match the username-based patterns {\tt /user/} and {\tt /profile?user=} during 2006-2013. 
By collecting approximately $\sim$ 3 million captures (2,997,954) for this period, we observe subscriber counts for approximately the one million top channels (1,047,839).

\subsubsection{Validation} We observe the highest-subscriber channels in our cohort and compare them to Wayback Machine captures of SocialBlade's top 500 lists for December 2010, 2011, and 2012.\footnote{Earliest list on the Wayback Machine is Dec. 2010. While top 5000 lists begin to appear during this time period, they are not frequent enough to align to our quarterly intervals.} 
We find that we are able to recreate SocialBlade's records with near-perfect accuracy, with our top 500~lists sharing $\sim$ 96\% of their users, largely in the same order.

\begin{table}[h!]
    \centering
    \begin{tabular}{cccc}
    \toprule
        Year & Overlap & $r$ & $\rho$\\
        \midrule
        Dec. 2010 & 485 / 500 & 0.999 & 0.995\\
        Dec. 2011 & 484 / 500 & 0.999 & 0.999\\
        Dec. 2012 & 479 / 500 & 0.999 & 0.981\\
    \bottomrule
    \end{tabular}
    \caption{Comparisons of the Social Blade top 500 channel lists and our top 500 subscriber lists for the same periods. Overlap is the number of channels shared in each top 500, while $r$ and $\rho$ are Pearson and Spearman correlations.}
    \label{cr_table}
\end{table}

\subsection{Cohort 2 (2014-2016)}
\subsubsection{Background and rationale}
We repeat the prior procedure for the time period from 2014 to 2016. 
We selected this time period because by 2014, YouTube had largely transitioned to the channel ID URL format, with channel links adopting the {\tt /channel/*} structure platform-wide. This transition created a period in which both username-based and ID-based URLs were archived in parallel on the Wayback Machine. Upon inspection, we find that username-format captures are substantially more prevalent, by factors of 7$\times$, 9$\times$, and 2$\times$ in 2014, 2015, and 2016, respectively (Figure~\ref{bars}; see also Table~\ref{time_table} in Appendix~\ref{app:captures}). Consequently, to maximize coverage during this period, we sample both username and channel ID formats for 2014-2016. Because archival coverage is substantially denser in these years, we sample in monthly resolution rather than quarterly (Figure~\ref{bars}).

\subsubsection{Collection}
We follow the practice outlined in Section~\ref{sec:data_1} of taking the most frequent channels and aim for $\sim$ 3 million captures distributed over the time period. Since usernames are more common than channel IDs, we split this between two million username captures and one million ID captures. This results in the top 248,330 usernames ($\sim$ 1.92 million pages) and the top 236,880 IDs ($\sim$ 939 thousand pages). 
This results in a total of 448,982 unique channels, with an average of 6.4 monthly captures per channel. 

\subsubsection{Validation}
In 2015, Social Blade began to release lists of the most popular channels (i.e., the most subscribed) within 16 content categories (e.g. Comedy, Gaming, Sports).\footnote{Channels on YouTube do not include a content category field, but videos do. To obtain channel categories, Social Blade collects the 10 most recent videos for each channel and selects the most common video category.}

To estimate coverage of \ours, we count how many of the top channels in each Social Blade list are present in the previous four months of our cohort.
In addition, we measure on these overlaps the Spearman rank correlation between a channel's rank in each Social Blade list (ordered by subscriber count) and its rank in \ours{} (also ordered by subscriber count). We also measure the Pearson correlation between the subscriber counts recorded by Social Blade and the subscriber counts recorded in \ours{} for the overlapping channels. These results are presented in Table \ref{categories}. 

\begin{table}[h!]
\centering
\small
\setlength{\tabcolsep}{2.25pt}

\begin{tabular}{r|cccccccc}
\toprule
& Show
& Pets
& NonPft
& Trvl
& Cars
& News
& Tech
& Sprt
\\
Subs & 16K & 29K & 29K & 31K & 77K & 102K & 157K & 170K\\ 
\midrule
08/15 \hspace{1pt} & 127 & 212 & 246 & 214 & 292 & 395 & 380 & 348 \\
12/15 \hspace{1pt} & 106 & 216 & 234 & 236 & 274 & 410 & 379 & 328 \\
04/16 \hspace{1pt} & 107 & 205 & 250 & 232 & 299 & 403 & 371 & 353 \\
08/16 \hspace{1pt} &  96 & 199 & 229 & 216 & 309 & 379 & 375 & 329\\

\toprule
& Edu
& Film
& Vlog
& DIY
& Comdy
& Gam
& Entr
& Mus
\\ 
Subs & 178K & 258K & 387K & 552K & 578K & 1.1M & 1.1M & 1.2M \\
\midrule
08/15 \hspace{1pt}  & 352 & 344 & 352 & 421 & 380 & 446 & 465 & 461 \\
12/15 \hspace{1pt} & 348 & 344 & 336 & 416 & 387 & 429 & 443 & 469 \\
04/16 \hspace{1pt} & 359 & 356 & 386 & 408 & 396 & 448 & 459 & 470 \\
08/16 \hspace{1pt} & 344 & 340 & 351 & 379 & 394 & 429 & 461 & 470 \\
\bottomrule
\end{tabular}
\caption{Overlap between the Social Blade top 500 channel lists per category and our top subscriber lists for the same periods. Categories are organized by median number of subscribers in August 2015. The Spearman rank correlation and Pearson correlation for subscriber counts are $\geq$ 0.99 in all cases.}
\label{categories}
\end{table}

Consistent with our earlier findings, the correlations remain near-perfect across categories. At the same time, as noted previously, coverage is skewed toward higher-subscriber channels, which has a disparate effect on less-subscribed categories. 

To provide a direct comparison to said earlier findings, we also compare the \textit{top 5000} overall lists on Social Blade to the \textit{top 5000} subscribed in \ours~ up to the same date. We observe that intersection between these monthly lists and our dataset remains around 90\% and still exhibits near perfect correlations.

\begin{table}[h!]
\centering
\small
\setlength{\tabcolsep}{3pt}
\renewcommand{\arraystretch}{1}
\resizebox{\linewidth}{!}{
\begin{tabular}{lccc|lccc}
\toprule
Month & Overlap & $\rho$ & $r$ & Month & Overlap & $\rho$ & $r$ \\
\midrule
Feb 2014 & 4278 & 0.999 & 1.0   & Jul 2015 & 4599 & 0.978 & 0.999 \\
Mar 2014 & 4582 & 0.996 & 1.0   & Aug 2015 & 4585 & 0.976 & 0.999 \\
Apr 2014 & 4655 & 0.995 & 1.0   & Oct 2015 & 4417 & 0.969 & 0.999 \\
May 2014  & 4689 & 0.995 & 1.0   & Nov 2015 & 4367 & 0.967 & 0.998 \\
Jun 2014 & 4714 & 0.990 & 1.0   & Dec 2015 & 4348 & 0.976 & 0.999 \\
Jul 2014 & 4713 & 0.993 & 1.0   & Jan 2016 & 4363 & 0.975 & 0.998 \\
Sep 2014 & 4721 & 0.991 & 0.999 & Feb 2016 & 4356 & 0.971 & 0.998 \\
Oct 2014 & 4768 & 0.993 & 1.0   & Mar 2016 & 4372 & 0.973 & 0.998 \\
Dec 2014 & 4718 & 0.987 & 1.0   & Apr 2016 & 4346 & 0.973 & 0.998 \\
Jan 2015 & 4720 & 0.986 & 0.999 & May 2016  & 4417 & 0.980 & 0.999 \\
Feb 2015 & 4678 & 0.983 & 0.999 & Jun 2016 & 4382 & 0.982 & 0.999 \\
Mar 2015 & 4687 & 0.985 & 1.0   & Jul 2016 & 4355 & 0.979 & 0.999 \\
Apr 2015 & 4673 & 0.983 & 0.999 & Sep 2016 & 4273 & 0.976 & 0.998 \\
May 2015  & 4652 & 0.984 & 0.999 & Nov 2016 & 4202 & 0.978 & 0.999 \\
Jun 2015 & 4627 & 0.983 & 0.999 & Dec 2016 & 4174 & 0.980 & 0.999 \\
\bottomrule \\
\end{tabular}}
\caption{Comparisons of the Social Blade top 5,000 channel lists and our top subscriber lists for the same periods.\tablefootnote{Monthly coverage over the period is incomplete due to insufficient Wayback Machine capturing.} Overlap is the number of channels that appear in both lists for a given period, while $r$ and $\rho$ are Pearson subscriber count correlation and the Spearman rank correlations, respectively.}
\label{5k}
\end{table}

\section{Who's on YouTube?} \label{sec:who}
\ours~offers a basic index of many creators on YouTube.
As we describe in our introduction, before this work, this simple yet fundamental resource for studying YouTube was not available for computational social research.
Because our work introduces this fundamental resource, we therefore  perform basic exploratory analysis to examine \textit{who} these creators are and the types of content they produce. 
As we explain in our introduction, it would not be possible to complete this basic yet foundational analysis in a replicable, transparent, and large scale way without \ours.

\begin{figure}[b!]
    \centering
\includegraphics[width=\linewidth]{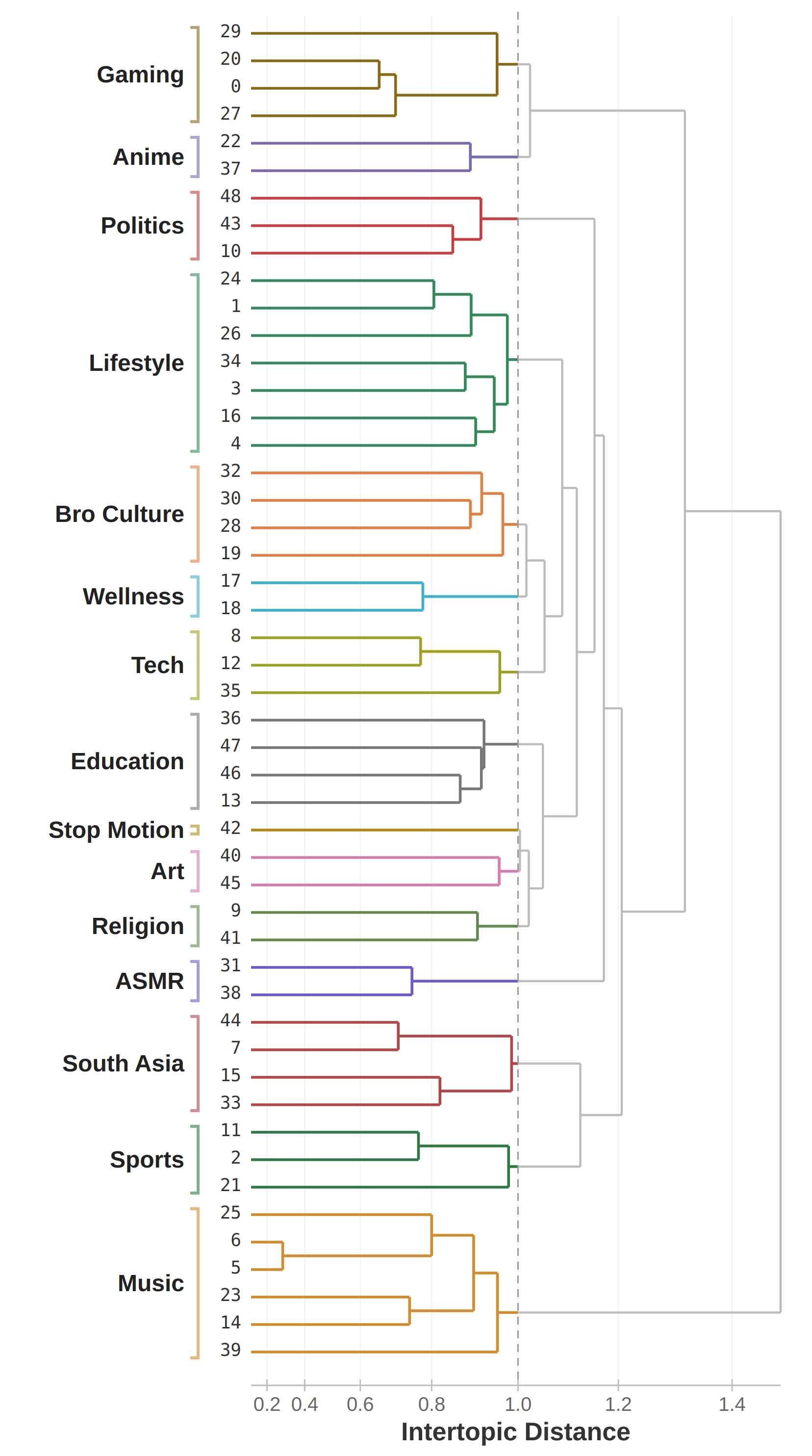}
    \caption{Dendrogram showing induced topics, labeled by 15 hierarchical topics. For the full 50 topics see App. \ref{app:topics}.}
    \label{dendrogram}
\end{figure}

For our analysis, we use the description fields of the 577,624 channel IDs identified in Section~\ref{s:creatorcoverage} as likely content creators. We filter these to only include English-language descriptions (using {\tt langid}) of at least 20 characters, yielding 208,893 channels, and fit a hierarchical topic model with BERTopic~\cite{grootendorst2022bertopic} to identify the themes and creator communities represented on the platform.\footnote{Although YouTube assigns each video to one of 16 official categories (Table~\ref{categories}), the labels are too coarse to capture the communities the channels serve or the production strategies they adopt.}

We visualize the resulting topics in the dendrogram in Figure~\ref{dendrogram}. The clusters correspond to distinct content genres, with color-coded branches highlighting the induced categories. The complete information for the topics extracted is included in Appendix~\ref{app:topics}. The diversity of genres encountered illustrates the breadth of creator communities represented on the platform and the range of cultural and functional niches that YouTube supports.

In addition, by linking these topics to the metadata of their respective channels, we uncover systematic variation across the inferred genres. For example, when examining channel creation dates, the most recent genres include VTubing, Political Campaigns, and Cryptocurrency, reflecting the rise of emerging technologies such as virtual reality and blockchain, as well as the cyclical nature of electoral activity. In contrast, the oldest genres correspond to long-standing institutions such as Universities, Museums, and Record Labels. Genres also differ in reach: Music channels attract the highest average views per video, while Local Politics (e.g., city council meetings) receive the fewest. In short, this analysis reveals different content creation strategies and audience dynamics across genres, highlighting a significant variation between groups in the YouTube creator ecosystem.
\section{Channels Grow in Predictable S-Curves} \label{sec:growth}

\ours~offers a way to observe the growth of YouTube channels over time.
We perform initial exploratory analysis of such growth, to help showcase the kinds of questions which researchers might examine with \ours~in future work.
 
\begin{figure}[h!]
    \centering
    \includegraphics[width=0.49\linewidth]{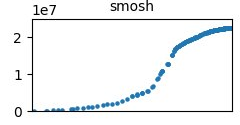}
    \includegraphics[width=0.49\linewidth]{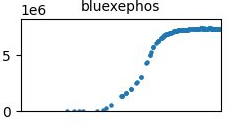}
    \caption{Examples of typical S-curve growth of YouTube channel subscribers through time.}
    \label{scurves}
\end{figure}

From this process, we observe that many channels exhibit logistic or S-curve growth, much like ecological populations \cite{verhulst1844recherches} or innovations in societies \cite{Rogers1964DiffusionOI}. 
Figure \ref{scurves} displays two examples drawn from the top 75 channels in January of 2014.\footnote{We show all 75 channels in the Appendix.}

Each channel exhibits the same pattern of subscriber counts that grows slowly, then quickly, and then eventually approach an asymptote.
This suggests that a YouTube channel may capture attention within a niche until it hits a fundamental limit (e.g., total potential audience size), much like a species might grow within an ecosystem until exhausting available resources.
Future research might examine unusual channels that expand their reach beyond their asymptote.
Research might also examine competitive dynamics between channels competing for attention within a niche, much like species competing for resources within an ecosystem. 

\begin{figure}[h!]
    \centering
    \includegraphics[width=110px]{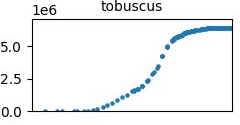}
    \includegraphics[width=110px]{smosh.png}\\[1em]
    \includegraphics[width=110px]{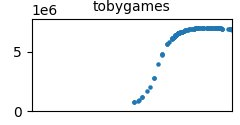}
    \includegraphics[width=110px]{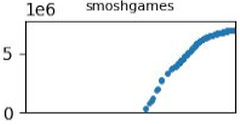}
    \caption{Examples of second channels overlaid under each other.}
    \label{sec}
\end{figure}

Yet while many channels exhibit S-curve growth, some channels deviate from this common pattern. 
In Figure \ref{sec}, we see that spin-off channels from already successful channels are able to skip the typical initial phase of slow growth.

\begin{figure}[h!]
    \centering
    \includegraphics[width=0.49\linewidth]{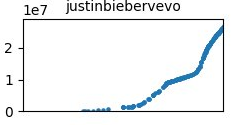}
    \includegraphics[width=0.49\linewidth]{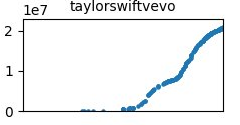}\\[1em]
    \includegraphics[width=0.49\linewidth]{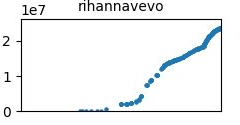}
    \includegraphics[width=0.49\linewidth]{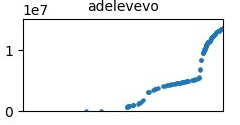}
    \caption{Examples of anomalous growth associated with celebrity musicians.}
    \label{vevo}
\end{figure}

Similarly, in Figure \ref{vevo}, we observe that some channels, such as celebrity musicians, grow in spurts that do not resemble the typical logistic curves. In each case, channels may be less reliant on YouTube platform dynamics to attract attention.
Closer analysis of both paradigmatic S-curve growth and S-curve outliers is a worthy topic for future research.

\section{Applications and Conclusion}~\label{sec:conclusion}

YouTube is core to contemporary mass media. 
In 2017, the platform announced that viewers around the world collectively watch one billion hours of YouTube every single day \cite{Goodrow2017}.
But despite the platform's core role in collecting, organizing, recommending, and distributing contemporary mass media, external researchers currently do not have any transparent, replicable, or large-scale way to observe basic information about the platform's creator economy, such as the set of creators on YouTube and their viewership through time.
The fundamental lack of visibility into a core piece of contemporary media infrastructure limits current understanding of the platform and its effects on audiences around the world.

For example, without an ability to observe YouTube creators and their audiences, researchers can not study if distributing incendiary online content or ``rage bait'' \cite{oup2025ragebait} helps creators build sustainable businesses on the platform, or if viewers eventually grow tired of creators who only offer sensational content.
Similarly, researchers can not study if YouTube channels that broadcast scientifically-vetted information grow more slowly or quickly than channels broadcasting unverified claims.
Moreover, researchers can not study how changes to the private platform (e.g., a 2012 algorithm change) affect what creators make and therefore audiences see on the popular platform.
Lack of basic visibility into the YouTube ecosystem impacts understanding of the modern world.

Our work begins to fill this fundamental gap with \ours, a large-scale, transparent, and easy-to-use dataset of YouTube channels and subscriber counts through time.
We construct \ours~by mining YouTube identifiers from Internet Archive captures across two decades of the YouTube ecosystem.
We then hide the complexities of this data gathering effort behind a simple API, implemented using a PIP package.
Because we build \ours~by synthesizing public channel page captures from the Internet Archive instead of through calls to a proprietary API, our steps are transparent and replicable, and do not rely on interaction with YouTube servers.
For these reasons, our approach might serve as an example for study of other social media ecosystems, which increasingly do not provide researchers access to platform data via HTTP requests or APIs.

\section*{Acknowledgements}

This work was partially supported by a Research \& Innovation Seed Grant from the Office of the Provost and Executive Vice Chancellor for Academic Affairs, Research \& Innovation Office (RIO) at the University of Colorado Boulder.

This work utilized the Alpine and Blanca high performance computing resources at the University of Colorado Boulder. Alpine is jointly funded by the University of Colorado Boulder, the University of Colorado Anschutz, Colorado State University, and the National Science Foundation (award 2201538). Blanca is jointly funded by computing users and the University of Colorado Boulder. Data storage was supported by the University of Colorado Boulder PetaLibrary.

\bibliography{0_aaai2026}

\appendix{ 
\section*{Paper Checklist}

\begin{enumerate}

\item For most authors...
\begin{enumerate}
    \item  Would answering this research question advance science without violating social contracts, such as violating privacy norms, perpetuating unfair profiling, exacerbating the socio-economic divide, or implying disrespect to societies or cultures?
    
    \answerYes{Yes.} 
  \item Do your main claims in the abstract and introduction accurately reflect the paper's contributions and scope?
    \answerYes{Yes.}
   \item Do you clarify how the proposed methodological approach is appropriate for the claims made? 
    \answerYes{Yes.}
   \item Do you clarify what are possible artifacts in the data used, given population-specific distributions?
    \answerYes{Yes. Sections \ref{sec:validation} and \ref{sec:data}}
  \item Did you describe the limitations of your work?
    \answerYes{Yes. Section \ref{sec:limitations}}
  \item Did you discuss any potential negative societal impacts of your work?
    \answerYes{Yes. Section \ref{sec:misuse}}
      \item Did you discuss any potential misuse of your work?
    \answerYes{Yes. Section \ref{sec:misuse}}
    \item Did you describe steps taken to prevent or mitigate potential negative outcomes of the research, such as data and model documentation, data anonymization, responsible release, access control, and the reproducibility of findings?
    \answerYes{Yes. Section \ref{sec:fair}}
  \item Have you read the ethics review guidelines and ensured that your paper conforms to them?
    \answerYes{Yes}
\end{enumerate}

\item Additionally, if your study involves hypotheses testing...
\begin{enumerate}
  \item Did you clearly state the assumptions underlying all theoretical results?
    \answerNA{N/A}
  \item Have you provided justifications for all theoretical results?
    \answerNA{N/A}
  \item Did you discuss competing hypotheses or theories that might challenge or complement your theoretical results?
    \answerNA{N/A}
  \item Have you considered alternative mechanisms or explanations that might account for the same outcomes observed in your study?
    \answerNA{N/A}
  \item Did you address potential biases or limitations in your theoretical framework?
    \answerNA{N/A}
  \item Have you related your theoretical results to the existing literature in social science?
    \answerNA{N/A}
  \item Did you discuss the implications of your theoretical results for policy, practice, or further research in the social science domain?
    \answerNA{N/A}
\end{enumerate}

\item Additionally, if you are including theoretical proofs...
\begin{enumerate}
  \item Did you state the full set of assumptions of all theoretical results?
    \answerNA{N/A}
	\item Did you include complete proofs of all theoretical results?
    \answerNA{N/A}
\end{enumerate}

\item Additionally, if you ran machine learning experiments...
\begin{enumerate}
  \item Did you include the code, data, and instructions needed to reproduce the main experimental results (either in the supplemental material or as a URL)?
    \answerNA{N/A}
  \item Did you specify all the training details (e.g., data splits, hyperparameters, how they were chosen)?
    \answerNA{N/A}
     \item Did you report error bars (e.g., with respect to the random seed after running experiments multiple times)?
    \answerNA{N/A}
	\item Did you include the total amount of compute and the type of resources used (e.g., type of GPUs, internal cluster, or cloud provider)?
    \answerNA{N/A}
     \item Do you justify how the proposed evaluation is sufficient and appropriate to the claims made? 
    \answerNA{N/A}
     \item Do you discuss what is ``the cost`` of misclassification and fault (in)tolerance?
    \answerNA{N/A}
  
\end{enumerate}

\item Additionally, if you are using existing assets (e.g., code, data, models) or curating/releasing new assets, \textbf{without compromising anonymity}...
\begin{enumerate}
  \item If your work uses existing assets, did you cite the creators?
    \answerYes{Yes}
  \item Did you mention the license of the assets?
    \answerYes{Yes. Section \ref{sec:fair}}
  \item Did you include any new assets in the supplemental material or as a URL?
    \answerYes{Yes}
  \item Did you discuss whether and how consent was obtained from people whose data you're using/curating?
    \answerNA{N/A}
  \item Did you discuss whether the data you are using/curating contains personally identifiable information or offensive content?
    \answerYes{Yes. Sections \ref{sec:fair} and 3.6}
\item If you are curating or releasing new datasets, did you discuss how you intend to make your datasets FAIR?
\answerYes{Yes. Section \ref{sec:fair}}
\item If you are curating or releasing new datasets, did you create a Datasheet for the Dataset? 
\answerYes{Yes. Appendix \ref{sec:datasheet}}
\end{enumerate}

\item Additionally, if you used crowdsourcing or conducted research with human subjects, \textbf{without compromising anonymity}...
\begin{enumerate}
  \item Did you include the full text of instructions given to participants and screenshots?
    \answerNA{N/A}
  \item Did you describe any potential participant risks, with mentions of Institutional Review Board (IRB) approvals?
    \answerNA{N/A}
  \item Did you include the estimated hourly wage paid to participants and the total amount spent on participant compensation?
    \answerNA{N/A}
   \item Did you discuss how data is stored, shared, and deidentified?
   \answerNA{N/A}
\end{enumerate}

\end{enumerate}
 \section{Channel Capture Counts}
\label{app:captures}
The counts of captures of each URL format across the dataset per year are shown in Tab.~\ref{time_table}
    
\begin{table*}[t]
\centering
\begin{tabular}{lccccccccc}
\toprule
\textbf{} & 2006 & 2007 & 2008 & 2009 & 2010 & 2011 & 2012 & 2013 & 2014 \\
\midrule
/user/ & 226 & 100k & 630k & 605k & 1.07M & 1.43M & 6.39M & 3.21M & 7.43M \\
/channel/ & 0 & 0 & 0 & 0 & 0 & 0 & 63.1K & 178K & 1.09M \\
/c/ & 0 & 0 & 0 & 0 & 0 & 0 & 0 & 0 & 13* \\
/@ & 0 & 0 & 0 & 0 & 0 & 0 & 0 & 0 & 0 \\
\toprule
 & 2015 & 2016 & 2017 & 2018 & 2019 & 2020 & 2021 & 2022 & 2023 \\
 \midrule
/user/ & 12.6M & 31.9M & 31.8M & 23.2M & 19.6M & 207M & 5.39M & 16.3M & 3.14M \\
/channel/ & 1.39M & 15.5M & 12.3M & 6.6M & 214M & 263M & 14.0M & 36.2M & 35.7M \\
/c/ & 6.87K & 164K & 77.6K & 248K & 5.65M & 43.4M & 2.09M & 7.42M & 1.11M \\
/@ & 0 & 0 & 22* & 19* & 5* & 46* & 59* & 6.22M & 88.2M \\
\bottomrule
\end{tabular}
\caption{The counts of captures of each URL format across the dataset's years. * indicates valid URLs uncovered before the announcement of a new format; in the case of handles, these were mainly vanity URLs (`/') that redirect to username.}
\label{time_table}
\end{table*}

\section{Induced Topics}\label{app:topics}

The resulting topics are visualized in the dendrogram in Figure~\ref{dendrogram}. The clusters correspond to distinct content genres, with color-coded branches highlighting the following categories from top to bottom: gaming, anime, politics, lifestyle, bro culture, wellness, tech, education, stop motion,  art, religion, ASMR, South Asia, sports, and music. These were generated using the default settings in BERTopic and then each of the genres were hand labeled based on their n-grams and exemplar documents. 

\begin{table*}[t]
\scriptsize
\centering
\resizebox{\textwidth}{!}{%
\begin{tabular}{llclrrrrr}
\toprule
\footnotesize Category & \footnotesize Topic ID & \footnotesize Top 4 tf-idf bigrams & \footnotesize Chnls &  \footnotesize Med. Subs & \footnotesize Med. Videos & \footnotesize Med. Views & \footnotesize Med. Join Date \\
\midrule
\multirow{4}{*}{Gaming}
 & 0 & ``minecraft'',``games'',``gaming'',``inquiries'' & 12381 & 7110.0 & 135.0 & 2759033.5 & 2014-11-02 \\
 & 20 & ``games'',``vr'',``indie'',``developer'' & 636 & 4030.0 & 96.0 & 1663895.5 & 2015-01-18 \\
 & 27 & ``games'',``guy'',``games channel'',``just guy'' & 428 & 454.0 & 52.0 & 253072.5 & 2013-09-13 \\
 & 29 & ``ram'',``cpu'',``specs'',``corsair'' & 404 & 25550.0 & 538.5 & 10274573.0 & 2013-09-11 \\
\midrule
\multirow{2}{*}{Anime}
 & 22 & ``anime'',``japan'',``japanese'',``manga'' & 524 & 22700.0 & 187.0 & 7098977.0 & 2015-03-30 \\
 & 37 & ``vtuber'',``hololive'',``vtubers'',``games'' & 275 & 19200.0 & 437.5 & 2826758.0 & 2019-09-08 \\
\midrule
\multirow{3}{*}{Politics}
 & 10 & ``district'',``congressional'',``running'',``candidate'' & 1331 & 53.0 & 25.0 & 11241.0 & 2019-07-16 \\
 & 43 & ``county'',``government'',``fairfax'',``montgomery'' & 178 & 975.0 & 492.0 & 246893.0 & 2013-09-21 \\
 & 48 & ``channel city'',``council'',``meetings'',``council meetings'' & 128 & 852.0 & 472.5 & 166826.0 & 2013-08-04 \\
\midrule
\multirow{7}{*}{Lifestyle}
 & 1 & ``recipes'',``cooking'',``makeup'',``cook'' & 4058 & 96000.0 & 305.0 & 15512933.0 & 2014-09-20 \\
 & 3 & ``car'',``cars'',``automotive'',``racing'' & 2082 & 46000.0 & 396.0 & 15323913.0 & 2014-01-09 \\
 & 4 & ``fuck'',``aint'',``yo'',``description'' & 1962 & 388.0 & 34.0 & 207574.0 & 2013-11-22 \\
 & 16 & ``rhymes'',``nursery'',``nursery rhymes'',``toys'' & 1021 & 359000.0 & 303.0 & 90983456.5 & 2015-06-04 \\
 & 24 & ``photography'',``painting'',``photoshop'',``tutorials'' & 505 & 75400.0 & 288.0 & 9059470.0 & 2014-01-14 \\
 & 26 & ``fishing'',``gardening'',``garden'',``farm'' & 478 & 90300.0 & 390.5 & 15986542.0 & 2014-08-04 \\
 & 34 & ``travel'',``destinations'',``hotels'',``vacation'' & 361 & 8610.0 & 206.0 & 2909292.5 & 2014-10-08 \\
\midrule
\multirow{4}{*}{\shortstack[l]{Bro Culture}}
 & 19 & ``wrestling'',``boxing'',``mma'',``martial'' & 681 & 71300.0 & 555.0 & 18477220.5 & 2014-11-10 \\
 & 28 & ``military'',``firearms'',``army'',``gun'' & 413 & 35700.0 & 268.0 & 7655933.0 & 2013-11-24 \\
 & 30 & ``politics'',``political'',``breaking'',``breaking news'' & 396 & 9265.0 & 913.0 & 3990193.0 & 2014-05-23 \\
 & 32 & ``ufo'',``paranormal'',``ufos'',``universe'' & 384 & 23300.0 & 239.0 & 3889416.0 & 2015-01-04 \\
\midrule
\multirow{2}{*}{Wellness}
 & 17 & ``fitness'',``workouts'',``workout'',``body'' & 925 & 72100.0 & 404.0 & 10461810.0 & 2014-05-15 \\
 & 18 & ``healthcare'',``medical'',``cancer'',``patients'' & 881 & 2595.0 & 215.0 & 586011.0 & 2013-01-10 \\
\midrule
\multirow{3}{*}{Tech}
 & 8 & ``trading'',``financial'',``crypto'',``blockchain'' & 1435 & 11000.0 & 288.0 & 1166695.0 & 2017-02-20 \\
 & 12 & ``solutions'',``cloud'',``security'',``software'' & 1237 & 2360.0 & 152.0 & 433151.5 & 2014-07-24 \\
 & 35 & ``wordpress'',``plugins'',``themes'',``plugin'' & 343 & 763.5 & 63.0 & 210890.5 & 2015-07-27 \\
\midrule
\multirow{4}{*}{Education}
 & 13 & ``faculty'',``academic'',``campus'',``undergraduate'' & 1227 & 2390.0 & 320.5 & 639590.5 & 2011-01-11 \\
 & 36 & ``grammar'',``spanish'',``lessons'',``vocabulary'' & 293 & 139500.0 & 312.0 & 11366256.0 & 2014-09-19 \\
 & 46 & ``exams'',``exam'',``preparation'',``ssc'' & 141 & 306000.0 & 2905.0 & 30549756.0 & 2016-04-11 \\
 & 47 & ``math'',``mathematics'',``maths'',``algebra'' & 128 & 66850.0 & 297.0 & 9321847.0 & 2013-07-30 \\
\midrule
Stop Motion & 42 & ``lego'',``stop motion'',``bricks'',``motion'' & 183 & 46500.0 & 186.0 & 13718228.0 & 2014-09-04 \\
\midrule
\multirow{2}{*}{Art}
 & 40 & ``museum'',``museums'',``exhibitions'',``contemporary'' & 198 & 2390.0 & 208.0 & 465558.0 & 2011-10-22 \\
 & 45 & ``orchestra'',``symphony'',``symphony orchestra'',``orchestras'' & 147 & 2310.0 & 312.0 & 1176978.0 & 2013-09-06 \\
\midrule
\multirow{2}{*}{Religion}
 & 9 & ``church'',``jesus'',``christ'',``bible'' & 1401 & 6950.0 & 485.0 & 973036.0 & 2014-08-05 \\
 & 41 & ``islam'',``islamic'',``quran'',``allah'' & 195 & 98700.0 & 487.0 & 12209886.0 & 2014-09-29 \\
\midrule
\multirow{2}{*}{ASMR}
 & 31 & ``sleep'',``relaxing'',``relaxation'',``meditation'' & 395 & 130000.0 & 252.0 & 26064041.0 & 2016-06-23 \\
 & 38 & ``asmr'',``asmr videos'',``sleep'',``relax'' & 259 & 156000.0 & 350.0 & 28293251.0 & 2016-11-12 \\
\midrule
\multirow{4}{*}{\shortstack[l]{South\\Asia}}
 & 7 & ``tamil'',``telugu'',``malayalam'',``bollywood'' & 1572 & 300000.0 & 2086.5 & 65760159.0 & 2015-11-28 \\
 & 15 & ``trailers'',``star wars'',``wars'',``movie trailers'' & 1048 & 10200.0 & 249.0 & 6310402.5 & 2014-07-04 \\
 & 33 & ``distribution'',``production company'',``distributor'',``cinema'' & 375 & 3880.0 & 137.5 & 1739211.5 & 2013-09-19 \\
 & 44 & ``songs'',``punjabi'',``devotional'',``haryanvi'' & 162 & 658500.0 & 381.0 & 200350455.0 & 2015-03-09 \\
\midrule
\multirow{3}{*}{Sports}
 & 2 & ``athletics'',``youtube page'',``music videos'',``athletics channel'' & 3786 & 29200.0 & 110.0 & 10343694.0 & 2012-12-17 \\
 & 11 & ``football'',``cricket'',``tennis'',``nba'' & 1259 & 39700.0 & 403.0 & 14435146.0 & 2015-07-07 \\
 & 21 & ``vevo'',``performances interviews'',``vevo official'',``official music'' & 528 & 134000.0 & 66.0 & 715579209.0 & 2009-12-11 \\
\midrule
\multirow{6}{*}{Music}
 & 5 & ``band'',``channel undefined'',``undefined'',``album'' & 1925 & 6980.0 & 286.0 & 14526057.0 & 2013-07-19 \\
 & 6 & ``album'',``debut'',``albums'',``awards'' & 1625 & 5390.0 & 110.0 & 4518404.0 & 2013-12-02 \\
 & 14 & ``guitar'',``songs'',``music channel'',``bands'' & 1206 & 14700.0 & 163.5 & 7214822.5 & 2012-12-07 \\
 & 23 & ``songwriter'',``singer'',``composer'',``music channel'' & 517 & 3035.0 & 45.0 & 840339.0 & 2012-09-04 \\
 & 25 & ``new album'',``album'',``preorder'',``new single'' & 489 & 59000.0 & 64.5 & 24825914.0 & 2012-09-13 \\
 & 39 & ``label'',``record label'',``records'',``independent record'' & 211 & 11200.0 & 231.0 & 5168271.0 & 2011-11-05 \\
\bottomrule
\end{tabular}}
\caption{Topic model of channel descriptions and the median metadata values for their corresponding channels.}
\label{topics}
\end{table*}
\begin{figure*}[t]
    \centering
    \includegraphics[width=0.9\linewidth]{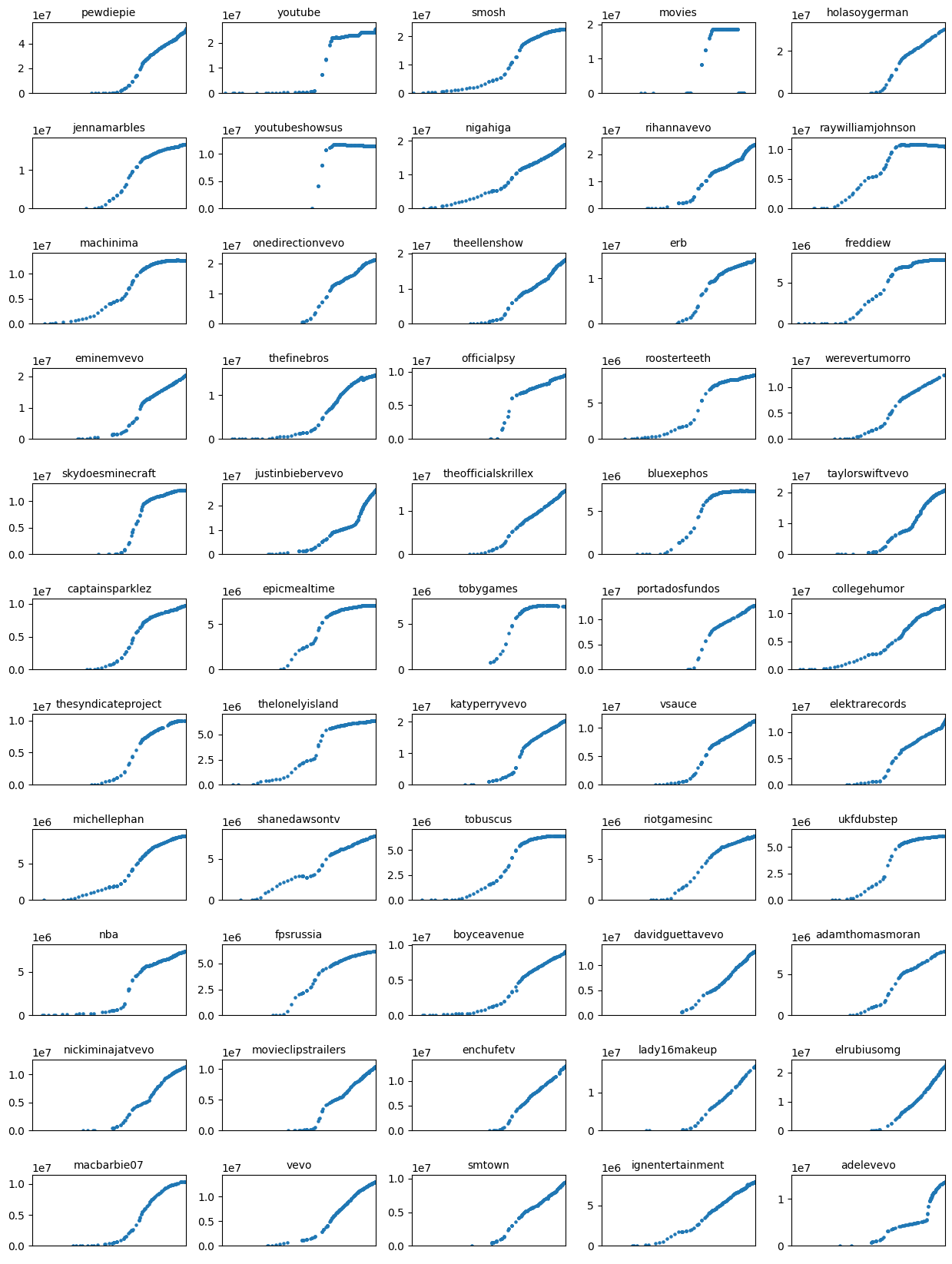}
    \caption{Channel growth (subscribers) over time (from 2006 through 2016) for top 75 most subscribed channels through 2013.}
    \label{growths}
\end{figure*}

\section{Datasheet for \ours{}}
\label{sec:datasheet}

\paragraph{Motivation}
\ours{} was created to enable transparent, replicable, and large-scale longitudinal research on the YouTube creator ecosystem. The official YouTube Data API does not provide a public census of channels and exposes only current (not historical) subscriber counts, limiting longitudinal measurement. \ours{} addresses this gap by releasing channel identifiers paired with time-stamped subscriber count observations.

\paragraph{Composition}
The dataset contains (i) YouTube channel identifiers and (ii) historical subscriber counts observed at specific points in time. The unit of analysis is a channel-level time series (repeated observations per channel). The dataset spans YouTube's history from 2006-2023 via archived snapshots. \ours{} does not include video-level data, content metadata, viewer-level data, comments, or sensitive personal information.

\paragraph{Collection process}
All records are derived from publicly accessible webpages archived by the Internet Archive's Wayback Machine, which stores time-stamped HTML snapshots of URLs over time. We extract subscriber counts from archived channel pages where present.

\paragraph{Preprocessing}
We normalize channel identifiers across URL formats when possible and align observations to fixed time intervals as described in the paper (e.g., quarterly or monthly sampling depending on archival density). Because YouTube's URL formats and page layout have changed over time, some channel attributes are inconsistently available; \ours{} focuses on subscriber counts, which are comparatively stable across archived captures.

\paragraph{Intended uses}
Appropriate uses include longitudinal analyses of channel growth and attention concentration, constructing cohorts of channels for downstream study, and research on measurement infrastructure under platform API restrictions.

\paragraph{Potential negative social impacts and misuse}
Although \ours{} releases only channel identifiers and subscriber counts, channel IDs can be used by third parties to link channels to additional platform data. This creates potential risks of profiling, targeting, or harassment of creators or communities. We mitigate these risks by releasing only the minimum information necessary for longitudinal measurement and explicitly discouraging harmful usages. 

\paragraph{Distribution and license}
The dataset is distributed via Zenodo and released under the Creative Commons Attribution 4.0 International (CC BY 4.0), which permits copying, redistribution, and adaptation of the dataset for any purpose (including commercial use), provided that users give appropriate credit, provide a link to the license, and indicate whether changes were made.

\paragraph{Maintenance and removal requests}
We may release updated versions via Zenodo versioning. If a channel owner requests removal, we will submit a deletion request to Zenodo and publish an updated version that tombstones the corresponding record.
}

\end{document}